\documentclass[aps,prb,twocolumn,showpacs]{revtex4}
\usepackage{amsmath}
\usepackage{amssymb}
\usepackage{epsfig}
\usepackage{revsymb}
\usepackage{ifthen} 

\newcommand{\scispc}{\ }
\newcommand{\scigroe}[3]{\ensuremath{#1\times10^{#2}\scispc\text{#3}}}
\newcommand{\magl}{\ensuremath{\ell}}
\newcommand{\energie}{\ensuremath{E}}
\newcommand{\pcfi}{\ensuremath{\varepsilon}}
\newcommand{\fillfac}{\ensuremath{\nu}}
\newcommand{\zyklfreq}{\ensuremath{\omega_c}}
\newcommand{\einhalb}{\ensuremath{\frac{1}{2}}}
\newcommand{\nplus}{\ensuremath{\text{n}^+}}
\newcommand{\Ugate}{\ensuremath{U_g}}
\newcommand{\etal}{\emph{et al.}}
\newcommand{\bild}[1]{\epsfig{file=fig#1,width=86mm}}
\newcommand{\reffig}[2][]{Fig.~\ref{fig#2}\ifthenelse{\equal{}{#1}}{}{(#1)}} 

\begin{document}
\title{Calculation and spectroscopy of the Landau band structure at a thin and
atomically precise tunneling barrier}

\author{Matthias Habl}\email{matthias.habl@physik.uni-r.de}
\author{Matthias Reinwald}
\author{Werner Wegscheider}
\affiliation{Universit\"at Regensburg, Institut f\"ur Experimentelle 
  und Angewandte Physik, 93040 Regensburg, Germany}

\author{Max Bichler}
\author{Gerhard Abstreiter}
\affiliation{Walter Schottky Institut, Technische
  Universit\"at M\"unchen, 85748 Garching, Germany}

\date{\today}

\begin{abstract}
Two laterally adjacent quantum Hall systems separated by an extended
barrier of a thickness on the order of the magnetic length possess a
complex Landau band structure in the vicinity of the line junction.
The energy dispersion is obtained from an exact quantum-mechanical
calculation of the single electron eigenstates for the coupled system
by representing the wave functions as a superposition of parabolic
cylinder functions. For orbit centers approaching the barrier, the
separation of two subsequent Landau levels is reduced from the
cyclotron energy to gaps which are much smaller. The position of the
anticrossings increases on the scale of the cyclotron energy as the
magnetic field is raised. In order to experimentally investigate a
particular gap at different field strengths but under constant filling
factor, a GaAs/AlGaAs heterostructure with a 52~\AA{} thick tunneling
barrier and a gate electrode for inducing the two-dimensional electron
systems was fabricated by the cleaved edge overgrowth method. The
shift of the gaps is observed as a displacement of the conductance
peaks on the scale of the filling factor. Besides this effect, which
is explained within the picture of Landau level mixing for an ideal
barrier, we report on signatures of quantum interferences at
imperfections of the barrier which act as tunneling centers. The main
features of the recent experiment of Yang, Kang \etal{} are reproduced
and discussed for different gate voltages. Quasiperiodic oscillations,
similar to the Aharonov Bohm effect at the quenched peak, are revealed
for low magnetic fields before the onset of the regular conductance
peaks.
\end{abstract}


\maketitle

\section{Introduction}
The bending of the Landau levels at the edge of a two-dimensional
electron system (2DES) is a well-investigated field, since it
represents an essential ingredient for the explanation of the quantum
Hall effect.\cite{Hal82,DS84,CSG92} Generally, one has been dealing with a
confining potential much wider than the magnetic length.  Only with
the technique of cleaved edge overgrowth (CEO)\cite{PWS90} it became
possible to define a sharp edge potential on the atomic
scale.\cite{HGR05} Kang, Yang and co-workers\cite{Kan00,Kan04,Kan05}
used this technique to separate a 2DES laterally by a
$88\scispc\text{\AA}$ wide and $212\scispc\text{meV}$ high
barrier. Using modulation doping, they achieved electron densities of
$1.1$ and $\scigroe{2.0}{11}{cm}^{-2}$. Variation of the magnetic
field and the bias voltage was employed to investigate the Landau band
structure.  Although the experiment of Kang \etal{} reveals the
principal features expected for the setup very well, there are some
points which are hardly explainable in the picture of Landau level
mixing. In particular, the first conductance peak for zero bias
occurs\cite{Kan00} at filling factor $\fillfac=1.2$ while the
dispersion predicts $\fillfac\approx4$. Furthermore, instead of the
observed broad conductance peaks, the small energy gaps in the single
electron picture would suggest very sharp peaks. The latter phenomenon
was discussed in several publications.\cite{MG01, KS01, TP00, ZS03,
KF03} The consideration of Coulomb interactions\cite{MG01, KS01} leads
to wider gaps which predict conductance peaks ($\Delta
\fillfac\approx0.01$) of still considerably smaller width than
detected in the experiment ($\Delta\fillfac\approx0.15$). While this
effect was also observed, we shall focus here on the dependence of the
peak position on the magnetic field. In addition to the regular
oscillations we shall investigate quasiperiodic structures of the
conductance at low magnetic fields and at the $\fillfac\approx2$ peak.

Section~\ref{sec:bandstructure} discusses the analytical solution of the
Schr\"o\-dinger equation for the coupled system which is outlined in
\reffig{1}. The wave functions are represented by parabolic cylinder
functions giving a convenient expression as a good starting point when
expanding the single electron model to an interacting model. With the
presented method, the dispersion can be determined for different sets
of parameters, especially for the strong coupling regime where the
picture of overlapping Landau bands as obtained at an infinitely high
barrier is not valid.

With our experiment (Sec.~\ref{sec:exp}) we want to get further
insight into the correspondence of the energy dispersion with the
conductance. Controlling the Fermi level allows us to investigate the
same gap at different magnetic fields. As a consequence, instead of
preparing samples with a thin AlGaAs layer of different thickness, the
magnetic field can be used to tune the effective shape of the barrier
while the filling factor is kept constant by means of the gate
electrode.  The advantage of providing variable electron densities
with \emph{one} sample is furthermore that small effects due to
unavoidable fluctuations between successive growth processes can be
excluded. This is essential when considering quantum
interferences\cite{Kan05,KF03} at tunneling centers caused by
imperfections of the barrier. With increasing electron density, a
shift of the conductance peaks towards higher filling factors is
observed which can be explained partly by the rise of the band gaps on
the scale of the cyclotron energy. While important properties like the
distortion of the $\fillfac\approx2$ peak by irregular large-period
features\cite{Kan05} are reproduced with our sample, there are
substantial differences compared to the modulation-doped structure of
Kang and co-workers.  The oscillations highly exceed the expected
limit by the conductance quantum, and the peaks are separated about
twice the distance on the scale of the filling factor.

\begin{figure}[t]
  \bild{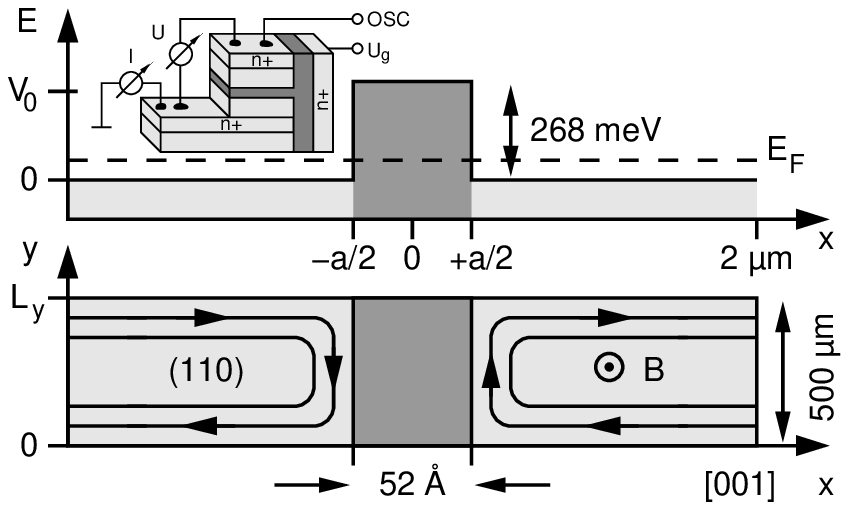}%
  \caption{\label{fig1}Shape of the conduction band edge and
    sketch of both 2DESs. The latter reside in two GaAs layers grown
    during the first MBE step on a (001) wafer. The tunneling barrier
    consists of about 18 monolayers of
    $\text{Al}_{0.34}\text{Ga}_{0.66}\text{As}$. The trajectories represent
    the two outermost edge channels. The inset depicts our sample structure.}
\end{figure}

\section{Band structure}\label{sec:bandstructure}
The goal of this section is to calculate the exact dispersion of a
single electron in the vicinity of the tunneling barrier by
determination of its eigenstates in the whole sample region composed
of both electron systems and the barrier. The spin is not considered
here. The time-independent Schr\"odinger equation $H\psi=E\psi$ in the
effective mass approximation,
\begin{equation}
  \left[\frac1{2m}\left(\mathbf{p}+e\mathbf{A}\right)^2
             +V(x)\right]\psi(x,y)=\energie\psi(x,y),
  \label{eqn:schroedinger1}
\end{equation}
has to be solved where $V(x)$ is the conduction band offset as shown
in \reffig{1}. For both materials the same effective mass of
$m=0.067\scispc m_e$ has been used. Neglecting all other edge
potentials, we are assuming infinitely extended electron systems in
the $x$-direction and periodic boundary conditions for the
$y$-direction. At first, the most general solution
of~\eqref{eqn:schroedinger1} for a region with constant $V(x)$ is
determined, and then, by applying the continuity conditions at the
barrier, the eigenenergies are calculated numerically.  The symmetry
of the system suggests the Landau gauge $\mathbf{A}=(0,xB,0)$ which
yields the Hamiltonian
\[
  H=\frac{\hbar^2}{2m}\left[\left(\frac1i\frac d{dy}+\frac x{\magl^2}\right)^2
    -\frac{d^2}{dx^2}\right]+V(x),
\]
where $\magl=\sqrt{\hbar/eB}$ is the magnetic length. The corresponding
Schr\"odinger equation is solved by the ansatz $\psi(x,y)\propto
e^{iky}\varphi(x)$.  The wave function is localized in the
$x$-direction and depicts a plain wave parallel to the barrier with
momentum $k\in\mathbb{Z}\times2\pi/L_y$. This results in the
one-dimensional differential equation
\[
  \frac{\hbar\zyklfreq}2\left[\left(k\magl+\frac x\magl\right)^2
  -\magl^2\frac{d^2}{dx^2}+\frac{2V(x)}{\hbar\zyklfreq}\right]\varphi(x)=
  \energie\varphi(x).
\]
In the following, the individual bands of the dispersion are
indexed by the Landau level number $n$. It is convenient to define the
dimensionless energy $\pcfi_{nk}$ by
\[
  \energie_{nk}=\left(\pcfi_{nk}+\textstyle\einhalb\right)\hbar\zyklfreq.
\]

\begin{figure}[t]
  \bild{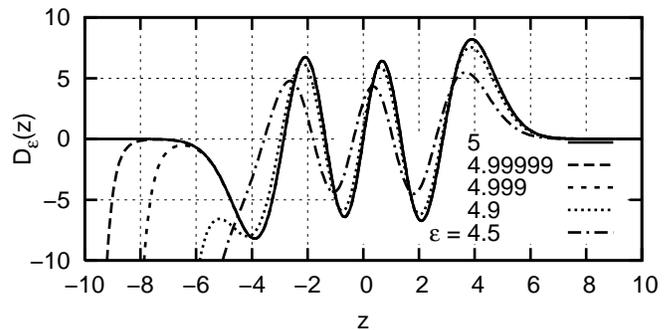}%
  \caption{\label{fig2}Parabolic cylinder function. When $\pcfi$
    approaches the integral value $n=5$, the divergence shifts to the
    left until it disappears completely.}
\end{figure}

\noindent The guiding center of the wave function is given by
$X=-k\magl^2=-\hbar k/eB$. The parameters $\xi=(x-X)/\magl$ and
$v=V/\hbar\zyklfreq$ are introduced in order to get an equation which
depends only on dimensionless variables and whose solution can be
found in the literature on higher transcendental
functions:\cite{Erd53, MOS66, Ros94}
\[
  \left[\frac{d^2}{d\xi^2}-\xi^2+2\left(\pcfi_{nk}-v(\xi)+\einhalb\right)\right]
  \varphi_{nk}(\xi)=0.
\]
Within regions of constant $v(\xi)$ the solution space is spanned
by two parabolic cylinder functions:
\[
  \varphi_{nk}(\xi)=\gamma_1D_{\pcfi_{nk}-v}(\xi\sqrt2)+
                    \gamma_2D_{\pcfi_{nk}-v}(-\xi\sqrt2).
\]
The function $D_\pcfi(z)$ is plotted in \reffig{2} for several
parameters.  At integer values of the energy, $\pcfi_{nk}=n$, the
parabolic cylinder function converges overall and it holds
$D_n(\xi\sqrt2) = 2^{-n/2}e^{-\xi^2/2}H_n(\xi)$.  This is the solution
of the harmonic oscillator which is valid for bulk states. Their
orbit center~$X$ coincides with the center of the wave function
and its distance to the barrier is at least about $(n+1)\magl$. For a
non-integer value of~$\pcfi$ the function $D_\pcfi(z)$ diverges when
its argument goes to minus infinity. Thus for the right 2DES
$\gamma_2$ has to be set to zero while the wave function in the left
electron system is $\varphi_{nk}(\xi)\propto
D_{\pcfi_{nk}}(-\xi\sqrt2)$.  The general solution of
\eqref{eqn:schroedinger1} is then given by
\begin{multline}
  \psi_{nk}(\xi,y)=e^{iky}\\
  \times\begin{cases}
    D_{\pcfi_{nk}}(-\xi\sqrt2)&\phantom|x\phantom|<-a/2\\
    D_{\pcfi_{nk}-v}(\xi\sqrt2)+\beta_{nk}D_{\pcfi_{nk}-v}
      (-\xi\sqrt2)&|x|\le\phantom-a/2\\
    D_{\pcfi_{nk}}(\xi\sqrt2)&\phantom|x\phantom|>-a/2.
  \end{cases}\label{eqn:wavefunction}
\end{multline}
For simplicity all prefactors have been dropped. The
characteristic parameters of the system are $V_0$, $a$, and $B$, but
the shape of the dispersion $\pcfi_n(X/\magl)$ is just determined by
the effective height~$v$ and width~$a/\magl$ of the barrier.

\begin{figure}[t]
  \bild{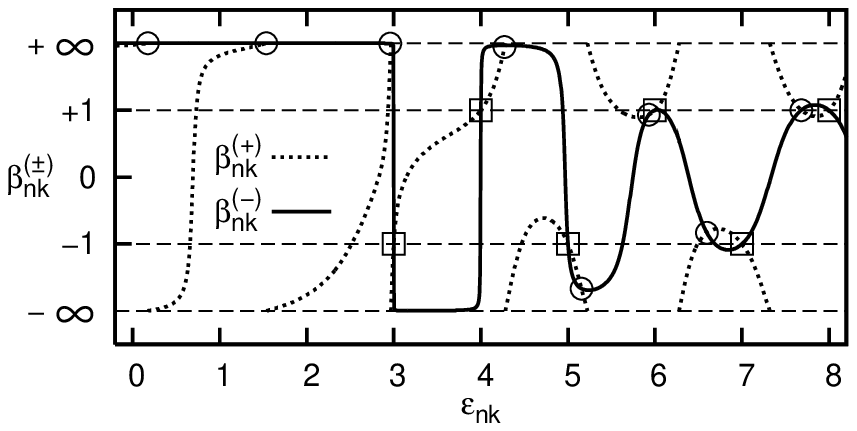}%
  \caption{\label{fig3}Two-dimensional parameter space
    $(\pcfi_n,\beta_n)$ for $B=9.2\scispc\text{T}$, the orbit center
    $X=-1.8\,\magl$, and a barrier of the height $v=3$ and width
    $a=179\scispc\text{\AA}=2.12\,\magl$. For each point on the dashed
    (solid) curve the boundary condition \eqref{eqn:boundcond} on the
    right (left) side of the barrier is fulfilled.  So the crossing
    points (\protect\raisebox{-.3ex}{\LARGE$\circ$}) represent the
    eigenstates of the system while pseudo solutions ($\square$)
    correspond to a vanishing denominator in
    \eqref{eqn:zaehlernenner}.}
\end{figure}

Both unknown variables, the energy~$\pcfi_{nk}$ and the mixing ratio
$\beta_{nk}$ of the parabolic cylinder functions inside the barrier,
are determined by the boundary conditions at the interfaces of the
barrier.  The continuity of the wave function and its derivative is
fulfilled when $d\,\log\varphi_{nk}(\xi)/d\xi$ is constant at the left
($-$) and right ($+$) side of the potential elevation:
\begin{gather}
  \left.\begin{gathered}
    \frac d{d\xi}\log D_{\pcfi_n}(\pm\xi\sqrt{2})=\\
    \frac d{d\xi}\log\left[D_{\pcfi_n-v}(-\xi\sqrt{2})+
        \beta_n D_{\pcfi_n-v}(\xi\sqrt{2})\right]
  \end{gathered}\right|_{\,\displaystyle\xi=\xi_\pm}\label{eqn:boundcond}\\
  \text{with}\quad\xi_\pm=\pm\frac a{2\magl}-\frac X{\magl}.\nonumber
\end{gather}
The evaluation of these expressions yields with\cite{Erd53}
$D^\prime_\pcfi(z)=\frac z2D_\pcfi(z)-D_{\pcfi+1}(z)$ the following
two equations:
\begin{gather}
  \begin{aligned}
    &\mp D_{\pcfi_n+1}(\pm\xi_\pm\sqrt{2})D_{\pcfi_n-v}(-\xi_\pm\sqrt{2})\\
    &\quad\mp\beta_nD_{\pcfi_n+1}(\pm\xi_\pm\sqrt{2})D_{\pcfi_n-v}(\xi_\pm\sqrt{2})\\
    &\quad-D_{\pcfi_n}(\pm\xi_\pm\sqrt{2})D_{\pcfi_n-v+1}(-\xi_\pm\sqrt{2})\\
    &\quad+\beta_nD_{\pcfi_n}(\pm\xi_\pm\sqrt{2})D_{\pcfi_n-v+1}(\xi_\pm\sqrt{2})
    \end{aligned}\quad\qquad\label{eqn:zaehlernenner}\\[-1.75ex]
  \frac{}{D_{\pcfi_n}(\pm\xi_\pm\sqrt{2})\Bigl(
   D_{\pcfi_n-v}(-\xi_\pm\sqrt{2})+\beta_nD_{\pcfi_n-v}(\xi_\pm\sqrt{2})\Bigr)}
  =0.\nonumber
\end{gather}
Finding \emph{all} solutions in the ranges $\pcfi_{nk}\in \left[0;
\pcfi_{\text{max}}\right]$ and $\beta_{nk}\in\left[-\infty;
+\infty\right]$ requires some effort beyond standard algorithms for
equation systems.  At fixed $X$ the numerators of
\eqref{eqn:zaehlernenner} are solved separately for $\beta_{nk}^{(+)}$
and $\beta_{nk}^{(-)}$ as shown in \reffig{3}. For $(\pcfi_{nk};\beta_{nk})=(v;-1)$,
$(v+1;1)$, $(v+2;-1)$, $\ldots$ both the numerators and denominators
vanish. Apart from these pseudo solutions all other crossings have to
be found to obtain the energy dispersion. This is intricate due to
partly very small singularities and high-valued regions of
$\beta_{nk}^{(\pm)}(\pcfi_{nk})$.

\begin{figure}[t]
  \bild{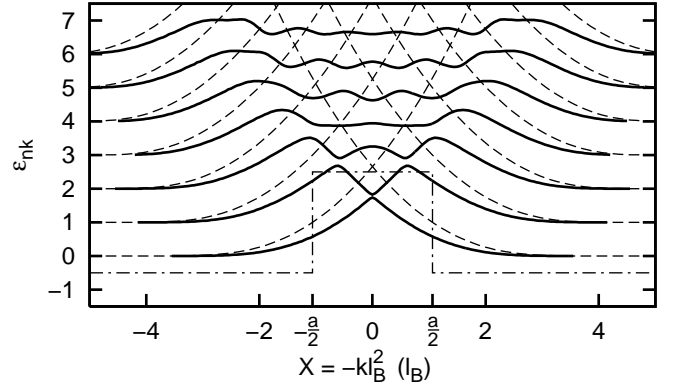}%
  \caption{\label{fig4}Dispersion relation (solid lines) for a low
    barrier. The parameters are the same as in \reffig{3}. For
    comparison, the eigenstates for an infinitely high barrier are
    plotted as dashed curves. The lowest gap inside the barrier is
    only $0.08\,\hbar\zyklfreq$ wide. The dash-pointed line denotes
    the barrier.}
\end{figure}

The results for the case of strongly coupled electron systems are
shown in \reffig{4} where the barrier has a height on the order of the
cyclotron energy.  Just like the wave functions themselves
(\reffig{2}), with subsequent index~$n$ an additional oscillation appears
in the band $\pcfi_{nk}$. While the gap between the two lowest Landau bands
is very small, it rises up to the order of $\hbar\zyklfreq$ for
eigenstates lying energetically above the barrier. The depicted
dispersion for an infinitely high barrier is obtained for the left
2DES by solving $D_{\pcfi_{nk}} \left(\sqrt2(\frac a2+X)
/\magl\right)=0$. The wave function is given by
\eqref{eqn:wavefunction} with a vanishing amplitude for $|x|\le a/2$.
A different ansatz in Ref.~\onlinecite{DS84} yields the same result.
The overlapping Landau bands can be used as a basis for further
perturbation calculations.  In contrast to the situation in
\reffig{4} this works well only when the appropriate Landau band
lies significantly below the barrier.

\begin{figure}[t]
  \bild{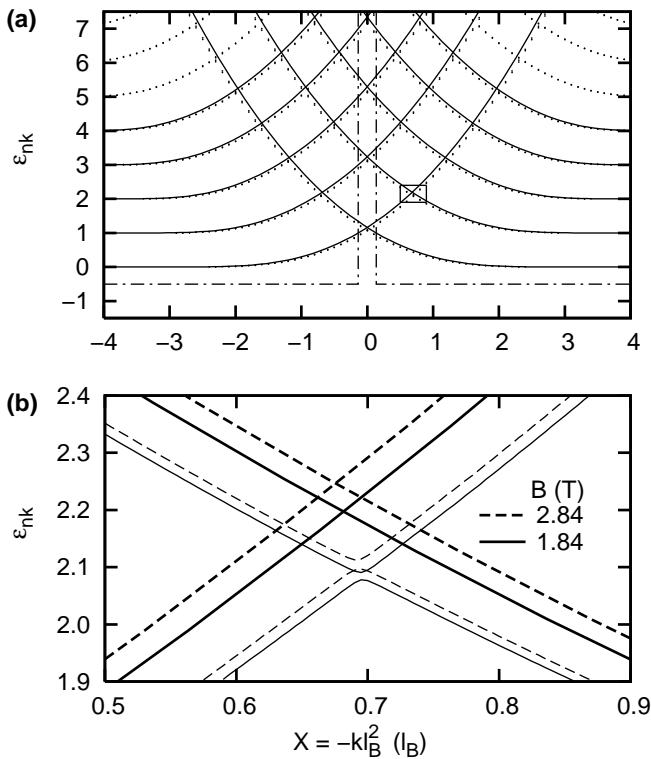}%
  \caption{\label{fig5}Band structure for our sample with
    $V_0=268\scispc\text{meV}$ and $a=52\scispc\text{\AA}$. (a)~The
    points are the result of the complete quantum-mechanical
    calculation, and the solid lines are corresponding to an infinitely
    high barrier.  The magnetic shift and the gaps are resolved by
    magnifying the marked rectangular region: (b)~Here, the bold lines
    stand for the overlapping Landau bands of two 2DESs separated by
    an infinitely high barrier. The crossing shifts by
    $\Delta\pcfi=0.05$ when the magnetic field increases as denoted.
    The anticrossing below (thin lines) divides the second and third
    Landau levels of the coupled system. With the change of
    $B$, the size of the gap rises from $0.013$ to
    $0.016\,\hbar\zyklfreq$ while its position shifts by
    $\Delta\pcfi=0.02$ (see also Table~\ref{tab1}).}
\end{figure}

\begin{figure}[t]
  \bild{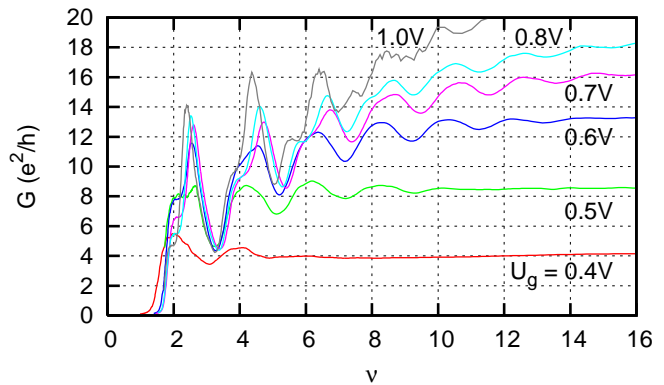}%
  \caption{\label{fig6}(Color online) Magneto-oscillations of the
    tunneling current at $400\scispc\text{mK}$. The conductance
    $G=I/V$ is plotted against the filling factor for several gate
    voltages $\Ugate$.}
\end{figure}
 
In order to suppress the bulk leakage current in our sample structure
(\reffig{1}), a rather high barrier with $V_0\gg\hbar\zyklfreq$
at typical field strengths was used. The corresponding band structure
of the weakly coupled system is shown in \reffig{5} for two
different magnetic fields. The deviation from the dispersion at an
infinitely high barrier is quite small, in contrast to the situation
in \reffig{4}.  But in detail, when looking at the dependence of
the gap positions on the magnetic field as demonstrated in
\reffig[b]{5}, the need for the outlined numerical solution is
apparent.  This predicts a shift of $\Delta\pcfi=0.02$ which amounts
about half the value than given by the superposition of two opposing
Landau bands, because for an increasing magnetic field not only the
rising effective barrier width $a/\magl\propto\sqrt B$ is considered
but also the compensation by the decrease of its effective height
$v\propto1/B$. For the same reason, at increasing magnetic field, the
coupling between the opposite edge channels is intensified which leads
to energy gaps rising faster than the cyclotron energy. The
investigation of the gap position on the scale of the Landau level
order is in the main focus of our experiment.

\section{Experiment}
\label{sec:exp} The heterostructure shown in \reffig{1} was fabricated in two
steps by employing the cleaved edge overgrowth method described in
detail in Ref.~\onlinecite{PWS90}. During a first growth process the
$a=52\scispc\text{\AA}$ thick tunneling barrier, two intrinsic regions
(each $2\scispc\mu\text{m}$), where the electron systems reside, and
two highly doped contact layers were produced. Immediately after
cleaving the sample \emph{in situ}, a gate structure with a
$d=100\scispc\text{nm}$ thick barrier and a $200\scispc\text{nm}$
\nplus-contact layer was grown on the freshly exposed (110) cleavage
plane. This design allows the variation of the electron density, but
the induced 2DESs cannot be contacted by a true four-point method.

While applying a magnetic field perpendicular to the cleavage plane,
the conductance was measured with lock-in technique using a
$10\scispc\text{nA}$ sinusoidal current at $17\scispc\text{Hz}$. In
\reffig{6} the conductance is plotted versus the filling factor
$\fillfac=nh/eB$ in order to allow direct comparison for different
electron densities. As the electron density~$n$ cannot be measured
directly, it is determined from the periodicity of the tunneling
resistance over $1/B$, see \reffig{7}. Takagaki and Ploog\cite{TP00}
already confirmed for this kind of tunneling spectroscopy an
equidistant conductance peak separation, namely, with a distance of
$\Delta \fillfac=2$ for a spin-degenerate system. Indeed, a constant
interval is also found experimentally at high accuracy as
demonstrated in \reffig{9}. In addition, the determined densities are
in agreement with by the capacitor model
\begin{equation}
  n(\Ugate)=\frac{\epsilon_r\epsilon_0}{de}\left(\Ugate-U_0\right)
  \label{eqn:capacitor}
\end{equation}
for the gate structure. Under low gate leakage, $n(\Ugate)$ fits
well to the experimental data when the dielectric constant\cite{Ada94}
$\epsilon_r=11.6$ for $\text{Al}_{0.34} \text{Ga}_{0.66} \text{As}$ at
$0\scispc\text{K}$ and the sample specific offset $U_0$ are used.  For
voltages $\agt0.6\scispc\text{V}$ an appreciable gate leakage
current emerges. The upper electron system depletes as the electron
supply from ground is restricted by the $52\scispc\text{\AA}$
tunneling barrier. The 2DESs are therefore shifted against each other
by an internal bias $U_2$ which is also plotted in \reffig{7}.

\begin{figure}[t]
  \bild{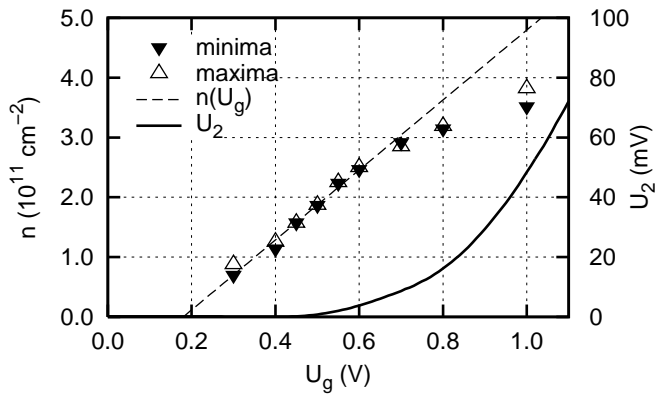}%
  \caption{\label{fig7}Electron density obtained by analyzing the
    resistance traces in the same way as is customary for
    Shubnikov-de~Haas oscillations. The density increases linearly
    until saturation effects appear at $\Ugate=0.7 \scispc
    \text{V}$. The relation \eqref{eqn:capacitor},
    $n=5.85\times10^{11}\,(U_g-0.18 \scispc \text{V})\scispc \text{cm}^{-2}/V$, is
    plotted with a fitted offset $U_0$ as the
    dashed line. The solid line represents the potential~$U_2$ of the
    upper contact layer (\reffig{1}).}
\end{figure}

In the picture of Landau level mixing, Kang and co-workers\cite{Kan00}
explained the oscillatory structure of the conductance by the
periodic appearance and disappearance of the inner counterpropagating
edge channels along the barrier (\reffig{1}) as the Fermi level is
varied. When the latter enters one of the small gaps, which divide
subsequent Landau bands, see \reffig[b]{5}, the corresponding edge
channels vanish while still remaining at the other edges of the
sample. By tunneling over the barrier, one large edge channel spanning
the whole sample is formed which gives rise to a conductance
peak.

\begin{figure}[t]
  \bild{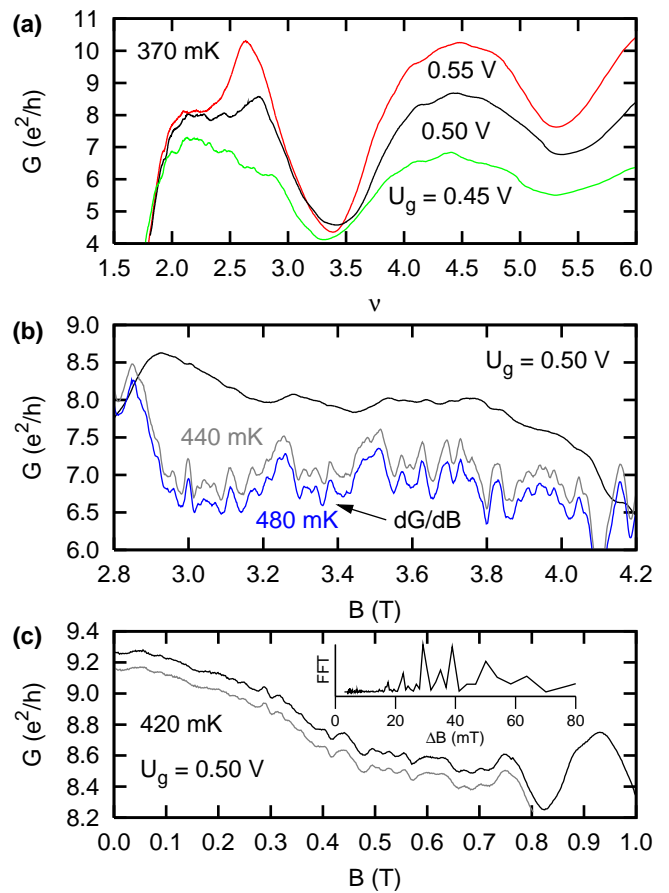}%
  \caption{\label{fig8}(Color online) Some features of the conductance
    traces. (a)~First two conductance peaks at different gate
    voltages. (b)~The magnification of the quenched peak around
    $\nu=2.5$---now plotted against the magnetic field---shows
    irregular oscillations, more clearly seen on its derivative which
    is reproduced for two subsequent magnetic field
    sweeps. (c)~Small-period oscillations at low magnetic fields are
    also highly reproducible at different field sweeps. The fast
    Fourier transform reveals about two distinct periods.}
\end{figure}

Although a four-point measurement was carried out, the interrupted
electron system is effectively contacted via two points, namely, by
two extended \nplus-layers. Consequently, in the quantum Hall regime
the conductance is expected to increase at most by two conductance
quanta when the tunneling of an edge channel is switched on. Several
edge channels may be involved at higher filling factors so that this
limit can be exceeded.\cite{TP00} In contrast to
Refs.~\onlinecite{Kan00, Kan04, Kan05}, where amplitudes on the order
of $0.1\scispc e^2/h$ are reported, we see conductance peaks with a
height of about $10\scispc e^2/h$, even at $\fillfac\approx 2$.  The
conductance varies from sample to sample, and when investigating a
sample which is half as wide, both the absolute conductance at
vanishing magnetic field and the amplitude of the magneto-oscillations
are reduced by the factor $1.8\pm0.2$ compared to a
$500\scispc\mu\text{m}$ sample obtained from the same growth. The
tunneling barrier faces to both highly doped contact layers which are
$250$ or $500\scispc\mu\text{m}$ long and reside at a distance of
$2\scispc\mu\text{m}$. Despite the extreme proportions of the 2DESs,
the enhancement of the conductance cannot be explained by
backscattering effects due to impurities since the magnetic length is
still much smaller than the width of the electron systems. It rather
seems that several macroscopic defects separate the active region
perpendicularly to the tunneling barrier into independent sections.
The formation of multiple edge channels in parallel overrides the
limit of $2\scispc e^2/h$. There is a couple of possible origins for
these defects. Small oval defects do not affect the lateral transport
of a (100)-2DES in the quantum Hall regime, but when randomly cleaved
during the CEO process, they massively distort the active region. The
same applies to low corrugations and steps arising when the sample
does not cleave perfectly, and partial damages of the unprotected
upper edge of the cleavage plane during processing are also
responsible for the separation.  A direct comparison of the
conductance amplitude with the values of Refs.~\onlinecite{Kan00,
Kan04, Kan05} is not reasonable as the number of independent sections
is not available and, in addition, the contact resistance and the
tunneling properties of the barrier had to be known for explaining any
differences. The barrier of Kang \etal{} was realized in a different
way, namely, as an $88\scispc\text{\AA}$ thick digital alloy of
$\text{Al}_{0.1} \text{Ga}_{0.9} \text{As}$/AlAs, and the contacts
were made by means of photolithography.

\begin{figure}[t]
  \bild{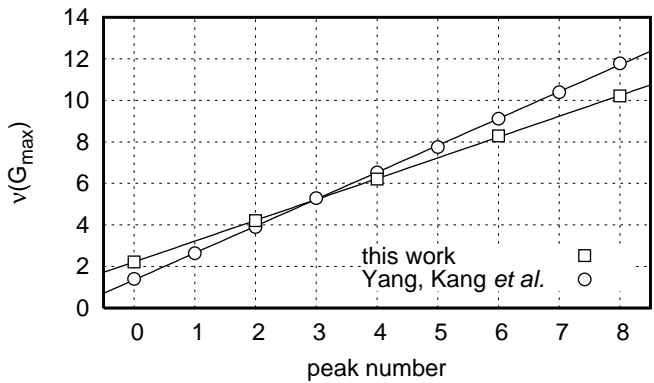}%
  \caption{\label{fig9}Position of the conductance peaks on the scale
    of the filling factor; $\square$, peaks shown in \reffig{6} at
    $\Ugate=0.5 \scispc \text{V}$ with $n= 1.87 \times
    10^{11}\scispc\text{cm}^{-2}$;
    \protect\raisebox{-.3ex}{\LARGE$\circ$}, data presented in
    Fig.~1a of Ref.~\onlinecite{Kan04}. The density $n = 1.89 \times
    10^{11} \scispc \text{cm}^{-2}$ is obtained from $R_{xx}(B)$ which
    is also plotted there. Straight lines: least mean square fit to
    the data points.}
\end{figure}

Even though our sample differs in considerable points, it displays an
important feature also discovered in the earlier
experiments.\cite{Kan05} For a non-biased system the leftmost peak in
\reffig[a]{8} is substantially cropped, and the peak at $\fillfac=4.5$
is also distorted. With increasing gate voltage, a smooth peak emerges
from the quenched peak at the left and the oscillatory structure of
the right peak becomes slightly weaker. While in
Ref.~\onlinecite{Kan05} a similar effect was reported for the
application of an external bias, in our sample the internal bias due
to gate leakage is responsible for the suppression of the irregular
oscillations: it increases in \reffig[a]{8} from $0.2$ to
$2\scispc\text{mV}$. The quenched peak, cf.~\reffig[b]{8}, consists of
a superposition of two different types of oscillatory features. The
first one has a large period of about $0.23\scispc\text{T}$ and can be
seen in $G(B)$ at rather high temperatures. The origin of these
oscillations remains unclear. However, small-period oscillations which
appear at dilution refrigerator temperatures were identified in
Ref.~\onlinecite{Kan05} as Aharonov-Bohm (AB) oscillations. They occur
due to strong coupling of the counterpropagating edge states via
imperfections in the barrier. Although the AB effect is
supposed\cite{Kan05} to be strongly suppressed above
$200\scispc\text{mK}$, we can reveal the oscillations still at
$480\scispc\text{mK}$ by averaging and differentiating as shown in
\reffig[b]{8}. They are quasiperiodic with not very distinct periods
in the range of $\Delta B=40\ldots70\scispc\text{mT}$. This
corresponds to a distance of $h/ea\Delta
B=7\ldots12\scispc\mu\text{m}$ between adjacent interference slits in
the barrier. Following from the average distance of the tunnel sites,
a lower bound of the phase-coherent length for the edge channels along
the barrier is estimated at $l_\Phi\approx20 \scispc\mu\text{m}$. The
AB oscillations cannot be detected for the $\fillfac\ge4.5$
peaks. Before the onset of the regular conductance peaks at about
$B=0.8\scispc\text{T}$, see \reffig[c]{8}, a series of small
quasiperiodic oscillations exists whose periods of
$30{-}50\scispc\text{mT}$ are independent of the field magnitude. With
increasing electron density, the emergence of these oscillations
decreases from $0.37\scispc\text{T}$ at $\Ugate=0.45\scispc\text{V}$
to $0.20 \scispc \text{T}$ at $\Ugate=0.55\scispc\text{V}$ as the
mobility rises due to improved screening. According to the relation
$B\mu_\Phi=2\pi$ the phase-coherent mobility is rated as
$\mu_\Phi\approx2 \times10^5\scispc\text{cm}^2/\text{Vs}$. The
irregular oscillations bear resemblance to the effect encountered at
the $\nu=2.5$ conductance peak, both according to the range of periods
and the good reproducibility which exists during the same cool-down
cycle.

The peaks feature a shoulder on the high field side as noticeable in
Figs. \ref{fig6} and \ref{fig8}(a). Although this bears resemblance
to the experiment of Huber \etal,\cite{HGR05} where the differential
tunnel conductance peak at filling factor $\fillfac^\perp=2.3$ has a
similar characteristic, we believe that this phenomenon can be
explained in the case of our experiment by the onset of
spin-splitting. The shoulder becomes more distinct with increasing
gate voltage as the mobility gets higher and the quenching due to
interferences at the tunneling centers is suppressed more
strongly. This feature is not contained in the data of Kang
\etal\cite{Kan00,Kan04} which, however, reveals a sequence of zero
bias conductance peaks as reproduced in \reffig{9} which are about
twice as dense. The strong suppression of spin-splitting in our sample
might be attributed to a rather low mobility which is indeterminable since
our design does not facilitate the characterization of an individual
electron system. Independently from the exact electron density, it
follows from \reffig{9} that the conductance peaks are spaced to a
great extent equally in respect of the filling factor.  The dispersion
in \reffig[a]{5} has a series of gaps with a uniform distance of
$\Delta\pcfi\sim1$ which are all shifted due to repulsion by the same
amount in relation to the bulk levels.  This expectation is confirmed
in principle by our data, with some modification discussed below
concerning the extent of the shift.  The peaks encountered in the
modulation-doped sample of Ref.~\onlinecite{Kan04} have a distance of
$\Delta\fillfac=1.29$ which means that there is no fixed relation with
respect to the bulk levels.

\begin{figure}[t]
  \bild{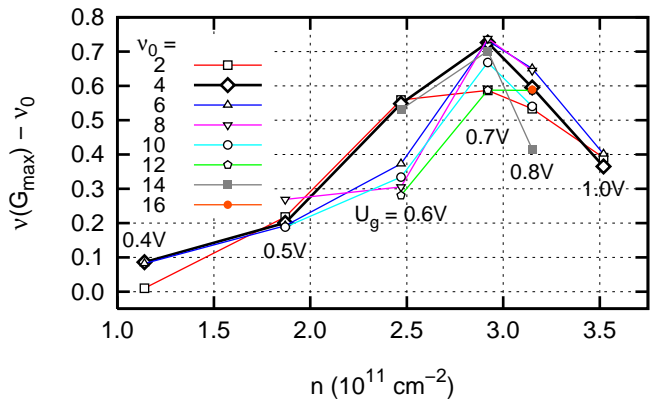}%
  \caption{\label{fig10}(Color online) Shift of conductance peaks
    versus electron density. For a fixed base filling factor $\fillfac_0$,
    the magnetic field rises from the left to the right.}
\end{figure}

\begin{table}[t]
  \caption{\label{tab1}Comparison of the filling factor~$\fillfac$ of the
    second conductance peak (\reffig{6}) and the lower ($\pcfi_1$)
    and upper ($\pcfi_2$) limit of the corresponding gap
    (\reffig[b]{5}) as a result of band structure calculations.}
  \begin{ruledtabular}
  \begin{tabular}{cccccc}
  \Ugate&$n$ $(10^{11}\scispc\text{cm}^{-2})$&
  $B$ $(\text{T})$&$\fillfac$&$\pcfi_1$&$\pcfi_2$\\
  \hline
  0.4&1.14&1.15&4.09&2.062&2.072\\
  0.5&1.87&1.84&4.20&2.078&2.091\\
  0.6&2.47&2.25&4.54&2.086&2.100\\
  0.7&2.92&2.56&4.73&2.092&2.107\\
  0.8&3.15&2.84&4.59&2.097&2.113
  \end{tabular}
  \end{ruledtabular}
\end{table}

The position of the conductance peaks in \reffig{6} depends on the
electron density. Initially, the peaks shift with increasing density
to higher filling factors. For $\Ugate>0.7\scispc\text{V}$ they shift
to lower values again. The displacement of the peak positions with
respect to the base filling factor $\fillfac_0=2$, $4$, $6$, $\ldots$
is depicted in \reffig{10}, and Table~\ref{tab1} summarizes the
properties of the conductance peak $\fillfac_0=4$.  All peaks evolve
in a very similar way. The decrease of the peak position at high gate
voltages can be explained\cite{Kan00} by the internal bias which is
caused by the leakage current trough the barrier.  The increase at low
electron densities is attributed to the rise of the gap position as
shown in \reffig[b]{5}. One would expect the Fermi energy to jump from
one bulk Landau level to another so that non-integer gap positions
cannot be realized at all. But as the peaks appear at
intermediate values, the Landau bands are assumed to be broadened
strongly. Then the relation $\fillfac=2\left(\pcfi+\einhalb\right)$
holds asymptotically for a spin-degenerate system. It allows to
compare the band structure with the conductance traces. The numerical
calculation yields for the second energy gap a position slightly above
the $n=2$ bulk Landau levels (\reffig{5}). Starting from
spin-degenerate eigenstates, a filling factor of at least
$\fillfac=5.2$ is needed to reach this gap, but the coincidence occurs
already at $\fillfac = 4.2\ldots4.6$. Obviously, the electron-electron
interaction between the opposite edge channels is not only responsible
for the increased gap, as discussed before, but also for its lowered
position. Section~\ref{sec:bandstructure} predicts a superlinear
increase of the gap position with increasing magnetic field. In
detail, the center of the second gap shifts by $\Delta\pcfi=0.018$
when the field increases from $1.15$ to $1.84\scispc\text{T}$ as
denoted in Table~\ref{tab1}. Hence, the corresponding conductance peak is
expected to shift by $\Delta\fillfac=0.036$, whereas the experiment
yields $\Delta\fillfac=0.11$. Another data set, shown in
\reffig[a]{8}, where the magnetic field is slightly shifted due to
different experimental conditions, exhibits a rise of
$\Delta\fillfac=0.06$ instead of $2\Delta\pcfi=0.026$. Consequently,
the actual gap position increases on the scale of the cyclotron energy
by the factor $2.7\pm0.4$ stronger than expected from the dispersion
calculated in the single electron approximation. This again points
towards many body effects which are not included in the calculation.

\section{Summary}
Our sample structure facilitates the investigation of a certain band
gap at different magnetic fields as the Fermi level can be adjusted by
means of the gate electrode. Apart from the exploitation of the field
effect, the heterostructure differs in the design of the contacts and
the barrier from the earlier experiments.\cite{Kan00, Kan04, Kan05}
Besides the main features of this kind of magneto-tunneling
spectroscopy we have especially reproduced at the $\fillfac_0=2$
conductance peak the prominent signature of quantum interferences
caused by imperfections in the barrier. At helium-3 temperatures the
peak is cropped and shows irregular large-period oscillations.  This
quenching can be partly suppressed by the internal bias due to gate
leakage. Weak AB oscillations are still detectable at
$480\scispc\text{mK}$. In addition, we found very similar
quasiperiodic oscillations at low magnetic fields before the onset of
the regular conductance peaks. In contrast to the results of Kang,
Yang \etal{}, our system is spin-degenerate and the conductance
exceeds the limit according to the Landauer-B\"uttiker formalism. The
latter effect is explained by the extended \nplus-contact layers in
conjunction with the separation of the active region into several
interrupted quantum Hall systems due to macroscopic defects at the
upper $[0\overline{1}0]$-ridge of the sample. Under low gate leakage,
measurements at different electron densities reveal a displacement of
the conductance peaks towards higher filling factors when the magnetic
field is increased. This is in accordance to the single electron band
structure which predicts an increase of the gap position on the scale
of the cyclotron energy. However, the expected shift is about $1/3$ of
$\Delta\fillfac$ encountered in the experiment.

Two different models have been developed hitherto for the
interpretation of this kind of experiment. The picture of Landau level
mixing\cite{Kan00, MG01, KS01, TP00} estimates the conductance peaks
as a consequence of vanishing counterpropagating edge channels when
the Fermi level coincides with a gap. For the purpose of investigating
the dependence of a certain gap on the magnetic field we have exactly
solved the Schr\"odinger equation in the Landau gauge for a single
electron which resides in a 2DES interrupted by a thin tunneling
barrier. Parabolic cylinder functions provide a compact representation
of the wave functions. The energy dispersion, determined by the
continuity at the heterojunctions, has gaps whose positions rise
faster than the cyclotron energy when the magnetic field increases. An
alternative picture for describing the physics at the junction is
based on the coupling of counterpropagating edge states via point
contacts in the barrier.\cite{Kan04, Kan05, KF03} The quasiperiodic
oscillations at low magnetic field and the properties of the quenched
conductance peak can rather be explained within this model.

\begin{acknowledgments}
M.~H. gratefully thanks Matthew Grayson for the useful discussion at
several occasions.  This work was supported by the \emph{Deutsche
Forschungsgemeinschaft} (DFG) via the priority program
\emph{Quanten-Hall-Systeme}.
\end{acknowledgments} 

\bibliography{habl}

\begin{thebibliography}{17}
\expandafter\ifx\csname natexlab\endcsname\relax\def\natexlab#1{#1}\fi
\expandafter\ifx\csname bibnamefont\endcsname\relax
  \def\bibnamefont#1{#1}\fi
\expandafter\ifx\csname bibfnamefont\endcsname\relax
  \def\bibfnamefont#1{#1}\fi
\expandafter\ifx\csname citenamefont\endcsname\relax
  \def\citenamefont#1{#1}\fi
\expandafter\ifx\csname url\endcsname\relax
  \def\url#1{\texttt{#1}}\fi
\expandafter\ifx\csname urlprefix\endcsname\relax\def\urlprefix{URL }\fi
\providecommand{\bibinfo}[2]{#2}
\providecommand{\eprint}[2][]{\url{#2}}

\bibitem[{\citenamefont{Halperin}(1982)}]{Hal82}
\bibinfo{author}{\bibfnamefont{B.~I.} \bibnamefont{Halperin}},
  \bibinfo{journal}{Phys.\ Rev.\ B} \textbf{\bibinfo{volume}{25}},
  \bibinfo{pages}{2185} (\bibinfo{year}{1982}).

\bibitem[{\citenamefont{MacDonald and St\v{r}eda}(1984)}]{DS84}
\bibinfo{author}{\bibfnamefont{A.~H.} \bibnamefont{MacDonald}}
  \bibnamefont{and}
  \bibinfo{author}{\bibfnamefont{P.}~\bibnamefont{St\v{r}eda}},
  \bibinfo{journal}{Phys.\ Rev.\ B} \textbf{\bibinfo{volume}{29}},
  \bibinfo{pages}{1616} (\bibinfo{year}{1984}).

\bibitem[{\citenamefont{Chklovskii et~al.}(1992)\citenamefont{Chklovskii,
  Shklovskii, and Glazman}}]{CSG92}
\bibinfo{author}{\bibfnamefont{D.~B.} \bibnamefont{Chklovskii}},
  \bibinfo{author}{\bibfnamefont{B.~I.} \bibnamefont{Shklovskii}},
  \bibnamefont{and} \bibinfo{author}{\bibfnamefont{L.~I.}
  \bibnamefont{Glazman}}, \bibinfo{journal}{Phys.\ Rev.\ B}
  \textbf{\bibinfo{volume}{46}}, \bibinfo{pages}{4026} (\bibinfo{year}{1992}).

\bibitem[{\citenamefont{Pfeiffer et~al.}(1990)\citenamefont{Pfeiffer, West,
  Stormer, Eisenstein, Baldwin, Gershoni, and Spector}}]{PWS90}
\bibinfo{author}{\bibfnamefont{L.}~\bibnamefont{Pfeiffer}},
  \bibinfo{author}{\bibfnamefont{K.~W.} \bibnamefont{West}},
  \bibinfo{author}{\bibfnamefont{H.~L.} \bibnamefont{Stormer}},
  \bibinfo{author}{\bibfnamefont{J.~P.} \bibnamefont{Eisenstein}},
  \bibinfo{author}{\bibfnamefont{K.~W.} \bibnamefont{Baldwin}},
  \bibinfo{author}{\bibfnamefont{D.}~\bibnamefont{Gershoni}}, \bibnamefont{and}
  \bibinfo{author}{\bibfnamefont{J.}~\bibnamefont{Spector}},
  \bibinfo{journal}{Appl.\ Phys.\ Lett.} \textbf{\bibinfo{volume}{56}},
  \bibinfo{pages}{1697} (\bibinfo{year}{1990}).

\bibitem[{\citenamefont{Huber et~al.}(2005)\citenamefont{Huber, Grayson,
  Rother, Biberacher, Wegscheider, and Abstreiter}}]{HGR05}
\bibinfo{author}{\bibfnamefont{M.}~\bibnamefont{Huber}},
  \bibinfo{author}{\bibfnamefont{M.}~\bibnamefont{Grayson}},
  \bibinfo{author}{\bibfnamefont{M.}~\bibnamefont{Rother}},
  \bibinfo{author}{\bibfnamefont{W.}~\bibnamefont{Biberacher}},
  \bibinfo{author}{\bibfnamefont{W.}~\bibnamefont{Wegscheider}},
  \bibnamefont{and}
  \bibinfo{author}{\bibfnamefont{G.}~\bibnamefont{Abstreiter}},
  \bibinfo{journal}{Phys.\ Rev.\ Lett.} \textbf{\bibinfo{volume}{94}},
  \bibinfo{pages}{016805} (\bibinfo{year}{2005}).

\bibitem[{\citenamefont{Kang et~al.}(2000)\citenamefont{Kang, Stormer,
  Pfeiffer, Baldwin, and West}}]{Kan00}
\bibinfo{author}{\bibfnamefont{W.}~\bibnamefont{Kang}},
  \bibinfo{author}{\bibfnamefont{H.~L.} \bibnamefont{Stormer}},
  \bibinfo{author}{\bibfnamefont{L.~N.} \bibnamefont{Pfeiffer}},
  \bibinfo{author}{\bibfnamefont{K.~W.} \bibnamefont{Baldwin}},
  \bibnamefont{and} \bibinfo{author}{\bibfnamefont{K.~W.} \bibnamefont{West}},
  \bibinfo{journal}{Nature} \textbf{\bibinfo{volume}{403}}, \bibinfo{pages}{59}
  (\bibinfo{year}{2000}).

\bibitem[{\citenamefont{Yang et~al.}(2004)\citenamefont{Yang, Kang, Baldwin,
  Pfeiffer, and West}}]{Kan04}
\bibinfo{author}{\bibfnamefont{I.}~\bibnamefont{Yang}},
  \bibinfo{author}{\bibfnamefont{W.}~\bibnamefont{Kang}},
  \bibinfo{author}{\bibfnamefont{K.~W.} \bibnamefont{Baldwin}},
  \bibinfo{author}{\bibfnamefont{L.~N.} \bibnamefont{Pfeiffer}},
  \bibnamefont{and} \bibinfo{author}{\bibfnamefont{K.~W.} \bibnamefont{West}},
  \bibinfo{journal}{Phys.\ Rev.\ Lett.} \textbf{\bibinfo{volume}{92}},
  \bibinfo{pages}{056802} (\bibinfo{year}{2004}).

\bibitem[{\citenamefont{Yang et~al.}(2005)\citenamefont{Yang, Kang, Pfeiffer,
  Baldwin, West, Kim, and Fradkin}}]{Kan05}
\bibinfo{author}{\bibfnamefont{I.}~\bibnamefont{Yang}},
  \bibinfo{author}{\bibfnamefont{W.}~\bibnamefont{Kang}},
  \bibinfo{author}{\bibfnamefont{L.~N.} \bibnamefont{Pfeiffer}},
  \bibinfo{author}{\bibfnamefont{K.~W.} \bibnamefont{Baldwin}},
  \bibinfo{author}{\bibfnamefont{K.~W.} \bibnamefont{West}},
  \bibinfo{author}{\bibfnamefont{E.-A.} \bibnamefont{Kim}}, \bibnamefont{and}
  \bibinfo{author}{\bibfnamefont{E.}~\bibnamefont{Fradkin}},
  \bibinfo{journal}{Phys.\ Rev.\ B} \textbf{\bibinfo{volume}{71}},
  \bibinfo{pages}{113312} (\bibinfo{year}{2005}).

\bibitem[{\citenamefont{Mitra and Girvin}(2001)}]{MG01}
\bibinfo{author}{\bibfnamefont{A.}~\bibnamefont{Mitra}} \bibnamefont{and}
  \bibinfo{author}{\bibfnamefont{S.~M.} \bibnamefont{Girvin}},
  \bibinfo{journal}{Phys.\ Rev.\ B} \textbf{\bibinfo{volume}{64}},
  \bibinfo{pages}{041309(R)} (\bibinfo{year}{2001}).

\bibitem[{\citenamefont{Kollar and Sachdev}(2002)}]{KS01}
\bibinfo{author}{\bibfnamefont{M.}~\bibnamefont{Kollar}} \bibnamefont{and}
  \bibinfo{author}{\bibfnamefont{S.}~\bibnamefont{Sachdev}},
  \bibinfo{journal}{Phys.\ Rev.\ B} \textbf{\bibinfo{volume}{65}},
  \bibinfo{pages}{121304(R)} (\bibinfo{year}{2002}).

\bibitem[{\citenamefont{Takagaki and Ploog}(2000)}]{TP00}
\bibinfo{author}{\bibfnamefont{Y.}~\bibnamefont{Takagaki}} \bibnamefont{and}
  \bibinfo{author}{\bibfnamefont{K.~H.} \bibnamefont{Ploog}},
  \bibinfo{journal}{Phys.\ Rev.\ B} \textbf{\bibinfo{volume}{62}},
  \bibinfo{pages}{3766} (\bibinfo{year}{2000}).

\bibitem[{\citenamefont{Z{\"u}licke and Shimshoni}(2003)}]{ZS03}
\bibinfo{author}{\bibfnamefont{U.}~\bibnamefont{Z{\"u}licke}} \bibnamefont{and}
  \bibinfo{author}{\bibfnamefont{E.}~\bibnamefont{Shimshoni}},
  \bibinfo{journal}{Phys.\ Rev.\ Lett.} \textbf{\bibinfo{volume}{90}},
  \bibinfo{pages}{026802} (\bibinfo{year}{2003}).

\bibitem[{\citenamefont{Kim and Fradkin}(2003)}]{KF03}
\bibinfo{author}{\bibfnamefont{E.-A.} \bibnamefont{Kim}} \bibnamefont{and}
  \bibinfo{author}{\bibfnamefont{E.}~\bibnamefont{Fradkin}},
  \bibinfo{journal}{Phys.\ Rev.\ B} \textbf{\bibinfo{volume}{67}},
  \bibinfo{pages}{045317} (\bibinfo{year}{2003}).

\bibitem[{\citenamefont{{Erd\'elyi} et~al.}(1953)\citenamefont{{Erd\'elyi},
  Magnus, Oberhettinger, and Tricomi}}]{Erd53}
\bibinfo{author}{\bibfnamefont{A.}~\bibnamefont{{Erd\'elyi}}},
  \bibinfo{author}{\bibfnamefont{W.}~\bibnamefont{Magnus}},
  \bibinfo{author}{\bibfnamefont{F.}~\bibnamefont{Oberhettinger}},
  \bibnamefont{and} \bibinfo{author}{\bibfnamefont{F.}~\bibnamefont{Tricomi}},
  \emph{\bibinfo{title}{Higher Transcendental Functions}}
  (\bibinfo{publisher}{McGraw-Hill}, \bibinfo{address}{New York},
  \bibinfo{year}{1953}).

\bibitem[{\citenamefont{Magnus et~al.}(1966)\citenamefont{Magnus,
  Oberhettinger, and Soni}}]{MOS66}
\bibinfo{author}{\bibfnamefont{W.}~\bibnamefont{Magnus}},
  \bibinfo{author}{\bibfnamefont{F.}~\bibnamefont{Oberhettinger}},
  \bibnamefont{and} \bibinfo{author}{\bibfnamefont{R.}~\bibnamefont{Soni}},
  \emph{\bibinfo{title}{Formulas and Theorems for the Special Functions of
  Mathmatical Physics}} (\bibinfo{publisher}{Springer},
  \bibinfo{address}{Berlin}, \bibinfo{year}{1966}).

\bibitem[{\citenamefont{{Barto\v{s}} and Rosenstein}(1994)}]{Ros94}
\bibinfo{author}{\bibfnamefont{I.}~\bibnamefont{{Barto\v{s}}}}
  \bibnamefont{and}
  \bibinfo{author}{\bibfnamefont{B.}~\bibnamefont{Rosenstein}},
  \bibinfo{journal}{J.\ Phys.~A} \textbf{\bibinfo{volume}{27}},
  \bibinfo{pages}{L53} (\bibinfo{year}{1994}).

\bibitem[{\citenamefont{Sadao}(1994)}]{Ada94}
\bibinfo{author}{\bibfnamefont{A.}~\bibnamefont{Sadao}},
  \emph{\bibinfo{title}{GaAs and related materials}} (\bibinfo{publisher}{World
  Scientific}, \bibinfo{address}{Singapore}, \bibinfo{year}{1994}).

\end{thebibliography}

\end{document}